\documentclass[aps,prl,a4,twocolumn,superscriptaddress,nofootinbib,showpacs,showkeys,preprintnumbers,]{revtex4-1}
\usepackage{graphicx}
\usepackage{dcolumn}
\usepackage{bm}
\usepackage[table]{xcolor}
\newcommand{\bef}{\begin{figure*}}
\newcommand{\eef}{\end{figure*}}

\newcommand{\be}{\begin{equation}}
\newcommand{\ee}{\end{equation}}
\newcommand{\bea}{\begin{eqnarray}}
\newcommand{\eea}{\end{eqnarray}}
\usepackage{lineno}

\begin{document}
\title{$K^{*0}$ meson production using a transport and a statistical hadronization model at energies covered by the RHIC beam energy scan}

\author{Aswini Kumar Sahoo}
\email{aswinis19@iiserbpr.ac.in}
\affiliation{Department of Physical Sciences, Indian Institute of Science Education and Research, Berhampur 760010, India}
\author{Md. Nasim}
\email{nasim@iiserbpr.ac.in}
\affiliation{Department of Physical Sciences, Indian Institute of Science Education and Research, Berhampur 760010, India}
\author{Subhash Singha}
\email{subhash@impcas.ac.cn}
\affiliation{Institute of Modern Physics, Chinese Academy of Sciences, Lanzhou, 73000, China}

\date{\today}
\begin{abstract}

In this paper, we discuss the centrality and energy dependence of $K^{*0}$ resonance production using ultrarelativistic quantum molecular dynamics (UrQMD) and thermal models. The $K^{*0}/K$ ratios obtained from the UrQMD and thermal models are compared with measurements done by the STAR experiment in Au+Au collisions at $\sqrt{s_{NN}}$ = 7.7, 11.5, 14.5, 19.6, 27, and 39 GeV.
The $K^{*0}/K$ ratio from the thermal model is consistent with data in most-peripheral collisions, however it overpredicts the ratio in central Au+Au collisions. This could be due to the fact that the thermal model does not have a hadronic rescattering phase, which is expected to be dominant in more central collisions. Furthermore, we have studied the $K^{*0}/K$ ratio from UrQMD by varying the hadron propagation time ($\tau$)  within the range 5 to 50 fm/c. It was found that the $K^{*0}/K$ ratio decreases with increasing $\tau$. Comparison between data and UrQMD suggest, one needs to consider a $\tau$ $\approx$ 10-50 fm/c to explain data at $\sqrt{s_{NN}}$ = 7.7-39 GeV in Au+Au collisions.  
We also predict the rapidity distribution of $K^{*0}$ from UrQMD which could be measured in the STAR beam energy scan phase II (BES-II) program.

\end{abstract}
\pacs{25.75.Ld}
\maketitle

\section{Introduction}

One of the major goals of heavy-ion collision is to study the properties of QCD matter produced in these collisions~\cite{star_white}. Just after the collision between two heavy-nuclei at relativistic speed, a deconfined state of quarks and gluons, commonly known as quark gluon plasma (QGP), is expected to be created~\cite{Bjorken}. Due to expansion, the temperature of the QGP decreases. When the temperature reaches the quark-hadron transition temperature, quarks and gluons are confined again to make hadrons. In the hadronic phase, particles can interact with themselves both elastically and inelastically. Chemical freeze-out happens when inelastic scattering between hadrons stops and kinetic or final freeze-out happens when particles do not interact among themselves, and elastic collision between the particles also ceases ~\cite{freeze_out_1,freeze_out_2,freeze_out_3}. After kinetic freeze-out, particles hit the detector. \\
Hadronic resonances can serve as unique probes to study the properties of hot QCD matter at different time scales, due to the different lifetimes of the different resonances~\cite{Brown_resonance, Markert_resonance}. For example, $K^{*0}(892)$ has lifetime $\approx$ $4.16$ fm/c~\cite{pdg}. Due to a short lifetime, $K^{*0}(892)$ mesons decay inside the fireball formed after the collision. The decay daughters of $K^{*0}(892)$ can undergo in-medium effects like rescattering and regeneration. For example, decay daughters of $K^{*0}(892)$ may undergo elastic scattering with other particles present in the medium. During the scattering process, momenta of daughter particles may get modified. Therefore, it may not be possible to reconstruct the parent. Hence this could cause a loss in the measured yield of $K^{*0}$. On the other hand, $\pi$ and $K$ mesons, present in the medium, can undergo pseudoelastic scattering~\cite{reco_issue_4} and form a $K^{*0}$ resonance between chemical and kinetic freeze-out. This is called regeneration. Due to regeneration, $K^{*0}$ yield is increased~\cite{reco_issue_1,reco_issue_2,reco_issue_3,reco_issue_4}. In order to have insight into these effects one can take the help of the resonance to non-resonance ratio (e.g., $K^{*0}/K$). If the rescattering process dominates, one naively expects the $K^{*0}/K$ ratio to decrease with increasing multiplicity. If the regeneration process dominates the ratio is expected to increase with the increasing multiplicity~\cite{star_kstar_2002, star_kstar_2005, star_kstar_2008, star_kstar_2011}. The loss or gain of resonance yield could depend on various factors, e.g., the lifetime of the hadronic phase, hadronic interaction cross sections of decay daughters, and particle density in the medium. Therefore, systematic study of the properties of  resonances like $K^{*0}$ may help us understand the effect of late-stage hadronic interactions.\\
 Previous measurements from STAR~\cite{star_kstar_2002, star_kstar_2005, star_kstar_2008, star_kstar_2011}, PHENIX~\cite{phenix_kstar_2014}, NA49~\cite{NA49_kstar_2011}, NA61~\cite{NA61_kstar_2020, NA61_kstar_2021}, ALICE~\cite{alice_kstar_2012,alice_kstar_2015,alice_kstar_2017,alice_kstar_2020_1,alice_kstar_2020_2,alice_kstar_2020_3,alice_kstar_2022} Collaborations show that rescattering effect can be the dominant mechanism in the late stage of hadronic medium produced in relativistic heavy-ion collisions. Various phenomenological studies also support this observation~\cite{reco_issue_4, pheno_kstar_2015, reco_issue_2}. 
Recently, the STAR Collaboration reported the measurement of  $K^{*0}$ production in Au+Au collisions at $\sqrt{s_{NN}}$ = 7.7, 11.5, 14.5, 19.6, 27 and 39 GeV~\cite{kstar_BES}. These data can be compared with ultrarelativistic quantum molecular dynamics (UrQMD) and thermal model~\cite{urqmd,thermus} calculations to get insight into the late stage hadronic medium produced at these energies. The thermal model has no hadronic phase, while UrQMD includes hadronic interaction among the particles. In UrQMD, one can vary the hadronic rescattering by varying the hadron propagation time. In this paper, the study is done by combining $K^{*0}$ and $\overline{K^{*0}}$ and is denoted by $K^{*0}$ in the text, unless otherwise specified. Also the charged kaons ($K^{\pm}$) are combined, denoted by $K$.\\
This paper is organized as follows. In Sec. II, we briefly discuss the thermal and UrQMD models. In Sec. III, we describe the study of  $K^{*0}$  at  RHIC beam energy scan (BES) phase I energies using the thermal and UrQMD model (version 3.3). A comparison with STAR data is shown. The results are summarized in Sec. IV.

\section{Model Description}
\subsection{The Thermal Model}
The  K$^{*0}$/K are obtained from statistical thermal model analyses of the produced particles using the THERMUS package~\cite{thermus} taking the grand-canonical ensemble (GCE).

In the GCE, for a hadron gas of volume $V$ and temperature $T$, the particle multiplicities are given by
\begin{eqnarray}
N_i^{GC} =
\frac{g_iV}{2\pi^2}  \sum_{k=1}^{\infty} (\mp1)^{k+1}
\frac{m_i^2T}{k}K_2 \left(\frac{km_i}{T} \right) \nonumber \\ e^{\beta k\mu_i},
\end{eqnarray}
where $K_2$ is the Bessel function of second order.  The chemical potential for particle species $i$ in this case is given by
\begin{eqnarray}
\mu_i = B_i \mu_B + Q_i \mu_Q + S_i \mu_S, 
\end{eqnarray}
where $B_i$, $S_i$, and $Q_i$ are the baryon number, strangeness, and
charge number, respectively, of hadron species $i$, and $\mu_B$,
$\mu_Q$, and $\mu_S$ are the respective chemical  potentials.

The freeze-out parameters ($T$, $\mu_B$, $\mu_Q$, and $\mu_S$) at different center-of-mass energies  are taken from the ref.~\cite{bulk_BES}. They are obtained by fitting the yields of $\pi^{\pm}$ , K$^{\pm}$,  $p$, $\bar{p}$, $\Lambda$, $\bar{\Lambda}$, $\Xi^{-}$, and  $\bar{\Xi}^{-}$ assuming the GCE. The freeze-out parameters are summarized in Table~\ref{tab-freeze-out-param}.


\bef
\begin{center}
\includegraphics[scale=0.7]{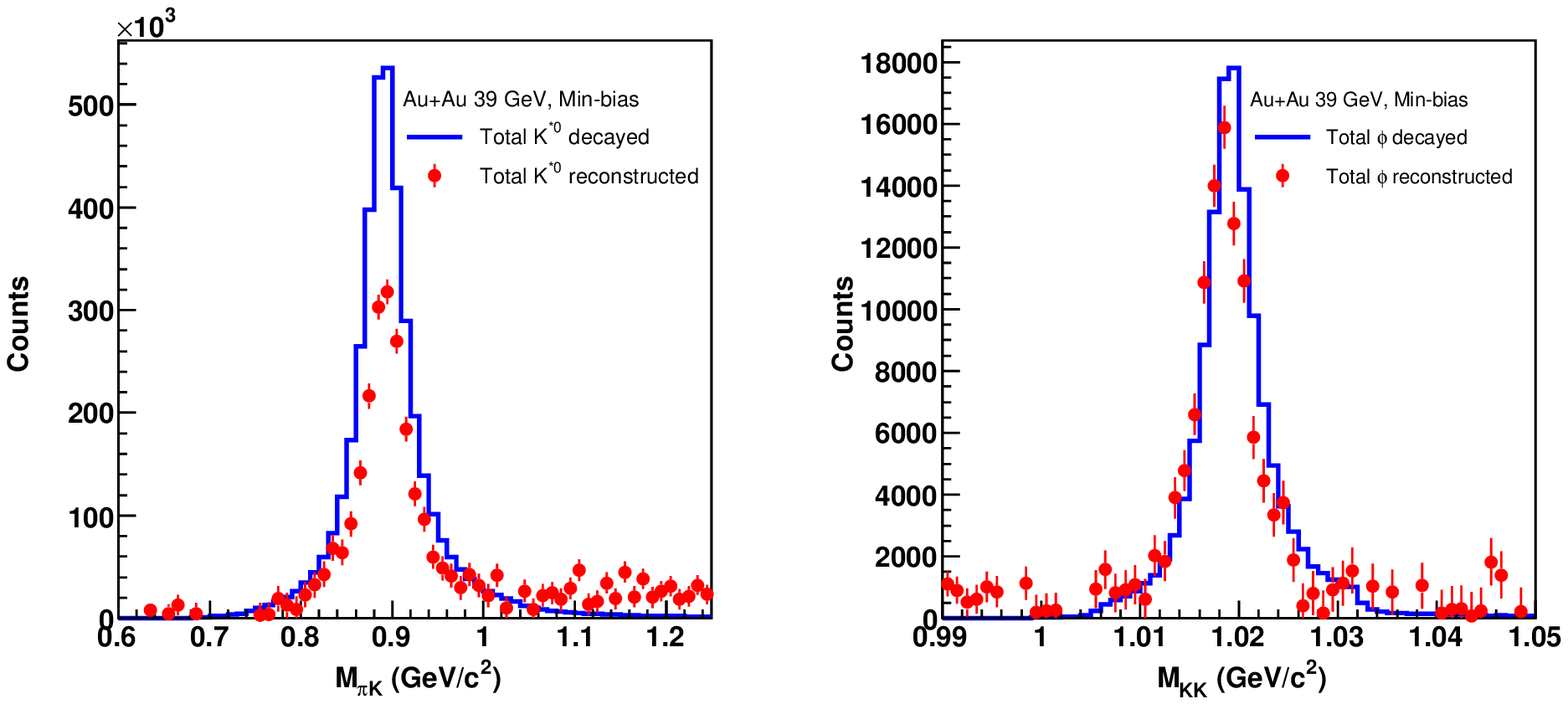}
\caption{(Color online) Invariant mass distribution of $K^{*0}$ ($K^{*0}\rightarrow K\pi$) and $\phi$ ($\phi\rightarrow KK$) for minimum-bias Au+Au collisions at 39 GeV and $\tau$= 20 fm/c. Here, the blue line denotes the total number of $K^{*0}$($\phi$) that have decayed and the red dots denote those which can be reconstructed. the errors on the data points are the statistical errors. }
\label{kstar_reco}
\end{center}
\eef

\begin{figure*}
\begin{center}
\includegraphics[scale=0.7]{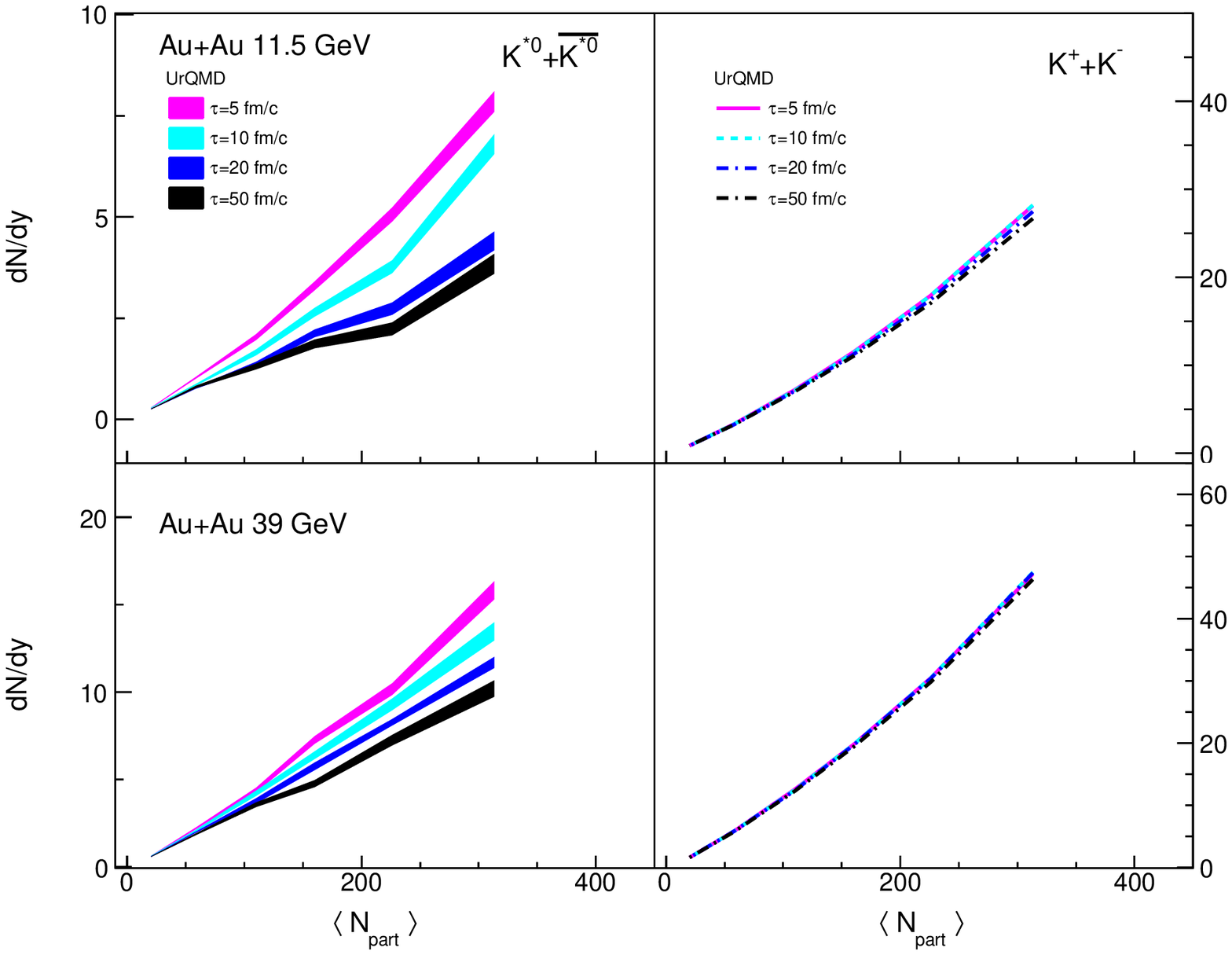}
\caption{(Color online) The $p_{T}$-integrated yield of $K^{*0}$ and charged kaons are measured using the UrQMD model for Au+Au collisions at 11.5 and 39 GeV}
\label{dndy_urqmd}
\end{center}
\end{figure*}

\begin{figure*}
\begin{center}
\includegraphics[scale=0.7]{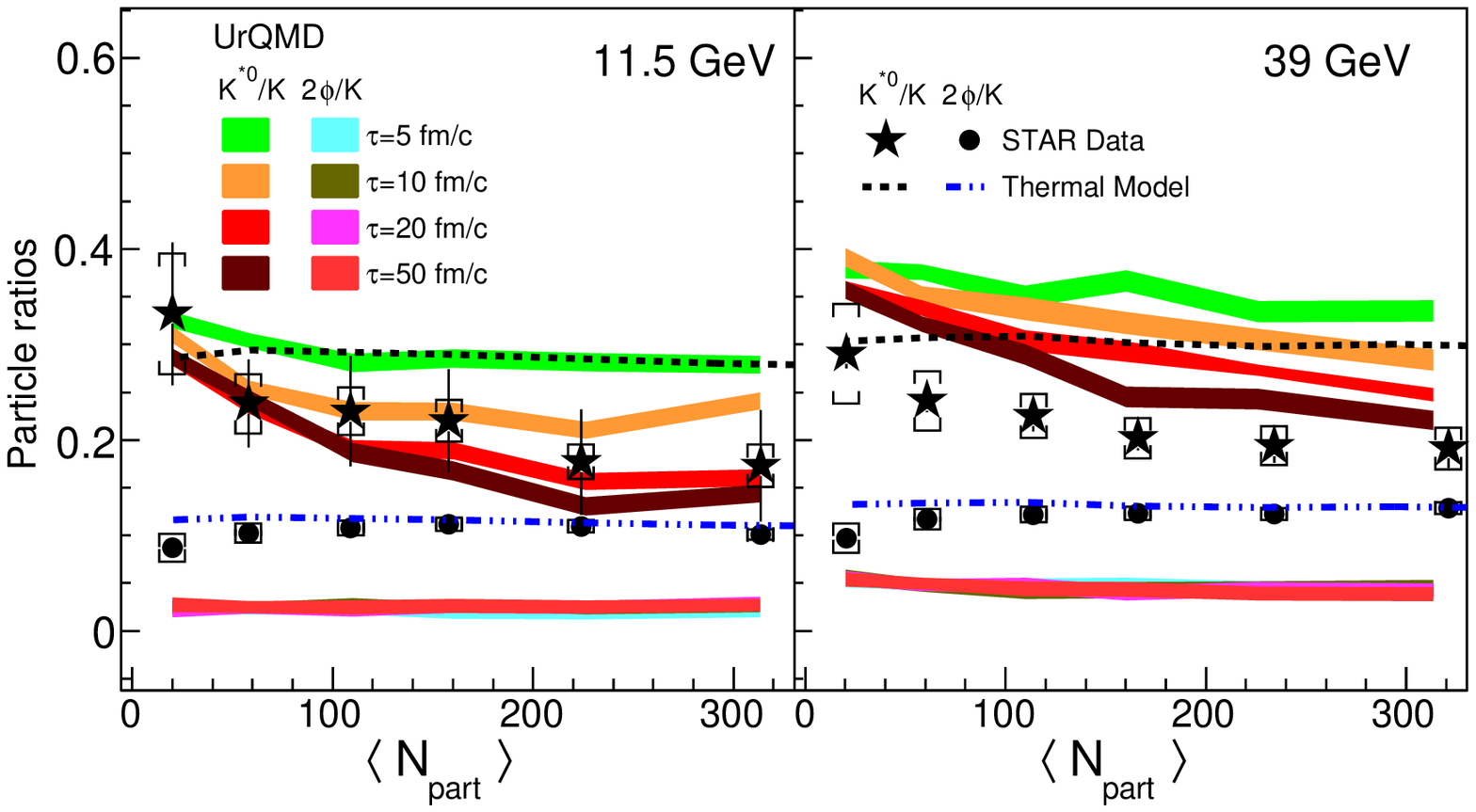}
\caption{(Color online) The $K^{*0}/K$ ratio vs $<N_{part}>$ measured at midrapidity from the STAR experiment~\cite{kstar_BES} compared with corresponding UrQMD model results at 11.5 and 39 GeV. The systematic and statistical uncertainties are shown by the caps and boxes on experimental data}
\label{ratio_npart}
\end{center}
\end{figure*}

\begin{figure*}
\begin{center}
\includegraphics[scale=0.7]{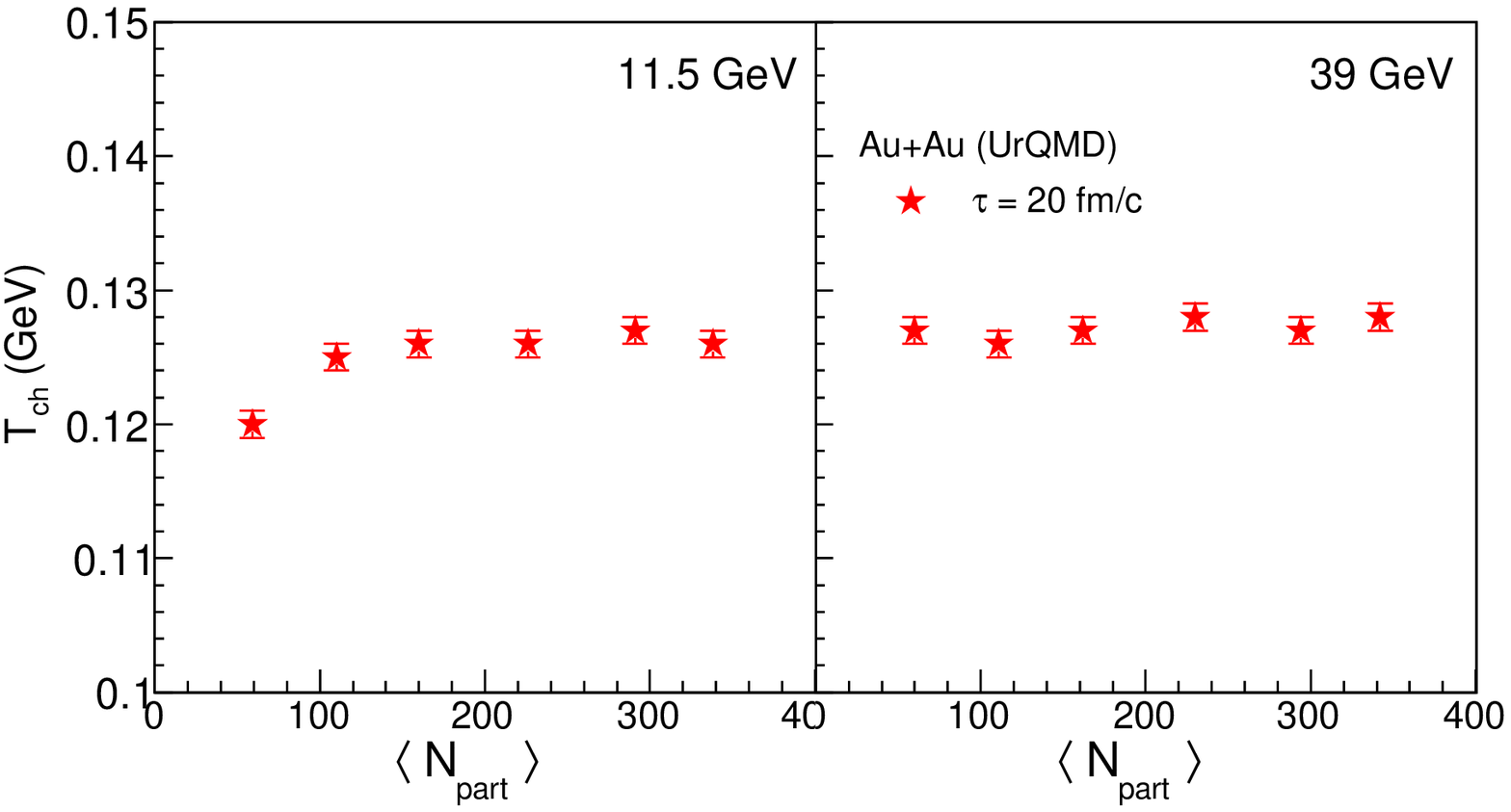}
\caption{(Color online) Chemical freeze-out temperature as  a function of $<N_{part}>$ for Au+Au collisions at 11.5 and 39 GeV extracted using the particle yield ratios from the UrQMD model.}
\label{tch_npart}
\end{center}
\end{figure*}

\subsection{The UrQMD Model}
The UrQMD (ultrarelativistic quantum  molecular dynamics) model~\cite{urqmd} is based on a microscopic transport theory where the
phase space descriptions of the reactions are important. It allows for the propagation
of all hadrons on classical trajectories in combination with stochastic binary 
scattering, color string formation, and resonance decay. It incorporates baryon-baryon,
meson-baryon, and meson-meson interactions; the collisional term includes more than 50 
baryon species and 45 meson species~\cite{urqmd}. UrQMD uses on-shell masses for all particles, i.e., the four-momentum vector of a particle satisfies its mass shell constraint. However, recent investigations show possible modification of properties of strange hadrons within a medium~\cite{modification_of_strangehadrons1,modification_of_strangehadrons2}.\\
 In the UrQMD model, one can vary the duration of the hadronic simulation ($\tau$) by tuning the \textit{"tim"} parameter in the input file~\cite{urqmd_link}. Our current approach, fixing $\tau$  to a particular value, essentially leads to an instant freeze-out. With increasing $\tau$ one allows the produced particles to interact among each other for a longer period. This could provide an opportunity to study the effect of hadronic re-scattering/regeneration on the yield of short-lived resonance particles, like $K^{*0}$. 

\section{ The reconstruction of resonances}
The hadronic resonances are short-lived particles, hence they are reconstructed from their respective decay daughters. In this analysis the $K^{*0}$ is reconstructed from its decay channel $K^{*0}(\overline{K^{*0}})\rightarrow K^{\pm}\pi^{\mp}$ with a branching ratio of 66.6\%~\cite{pdg}. The reconstruction of the resonance is done in the same way as it is done in experiment~\cite{kstar_BES}. First the $K^{*0}$ is reconstructed by combining all the $K\pi$ pairs in a same event via the invariant mass method. The uncorrelated pairs, known as combinatorial background are estimated using the pair-rotation method. The signal is obtained by subtracting the background from the same event $K\pi$ pairs. 
The particle yield is calculated by integrating the invariant mass spectra within 3$\sigma$ (width of the resonance) from its mean position and scaled by the branching ratio. 
Note that higher mass resonances [e.g. $f_0$(980)] may contribute to the yield, especially towards the tail of the invariant mass distribution. We extracted the particle yield by integrating the invariant mass within 3$\sigma$ from its mass peak position, where the contribution from higher mass resonances is expected to be negligible. In a future analysis, a detailed study on the change in line shape due to higher mass resonances can be performed by tagging the parent resonance and its daughters.\\
Figure~\ref{kstar_reco} shows the invariant mass distribution of $K^{*0}(\overline{K^{*0}})\rightarrow K^{\pm}\pi^{\mp}$. The blue line denotes the invariant mass distribution of $K^{*0}$ that have actually decayed. This information is available from the collision history file ({\it test.f15}) of UrQMD. which follows a nice Breit-Wigner function. The red data points denote the invariant mass distribution of the $K^{*0}$ that is reconstructed from the final freeze-out particle information contained in file {\it test.f14}. The reconstructed $K^{*0}$ count is less than the actually decayed ones, because the decay daughters experienced in-medium interactions, which makes the parent resonance unreconstructable. Since $\phi$ has a longer lifetime, its decay daughters remains unaffected by the medium and the number of reconstructed $\phi$ mesons is consistent with the actual $\phi$- meson decays.\\
We have fitted the $K^{*0}$ and  $\phi$ meson invariant mass distribution with a Breit-Wigner function. The resulting  $K^{*0}$ mass and width align closely with the corresponding values reported by the Particle Data Group (PDG). For $\phi$ mesons, we observe a change  $\approx$ 1-2 $\%$ in reconstructed mass and width  compared to PDG values. This might be the effect of in-medium interaction, as reported in previous studies~\cite{broadening_of_width1,broadening_of_width2}. 
\section{Results}
\subsection{Yield of $K^{*0}$ and charged kaons ($K^{\pm}$) from the UrQMD model}
 Figure~\ref{dndy_urqmd} shows yields (dN/dy) of $K^{*0}$ and charged kaon ($K^{\pm}$) as a function of number of participating nucleons ($N_{part}$). The measurements are done at midrapidity ($|y| < 1.0$ for $K^{*0}$ and $|y| < 0.1$ for kaons) in Au+Au collisions at $\sqrt{s_{NN}}$ = 11.5 and 39 GeV in order to be consistent with published results from STAR~\cite{kstar_BES, bulk_BES}. The results are obtained by varying $\tau$ from 5 to 50 fm/c for all STAR BES energies from 7.7 to 39 GeV. \\ 
 Figure~\ref{dndy_urqmd} shows that the centrality dependence of charged kaon yield remains independent of $\tau$, where as the $K^{*0}$ yield decreases with increase in $\tau$. The decrease in $K^{*0}$ yield is due to the rescattering of daughter particles in the hadronic phase, which is included in UrQMD.  
 




\bef
\begin{center}
\includegraphics[scale=0.7]{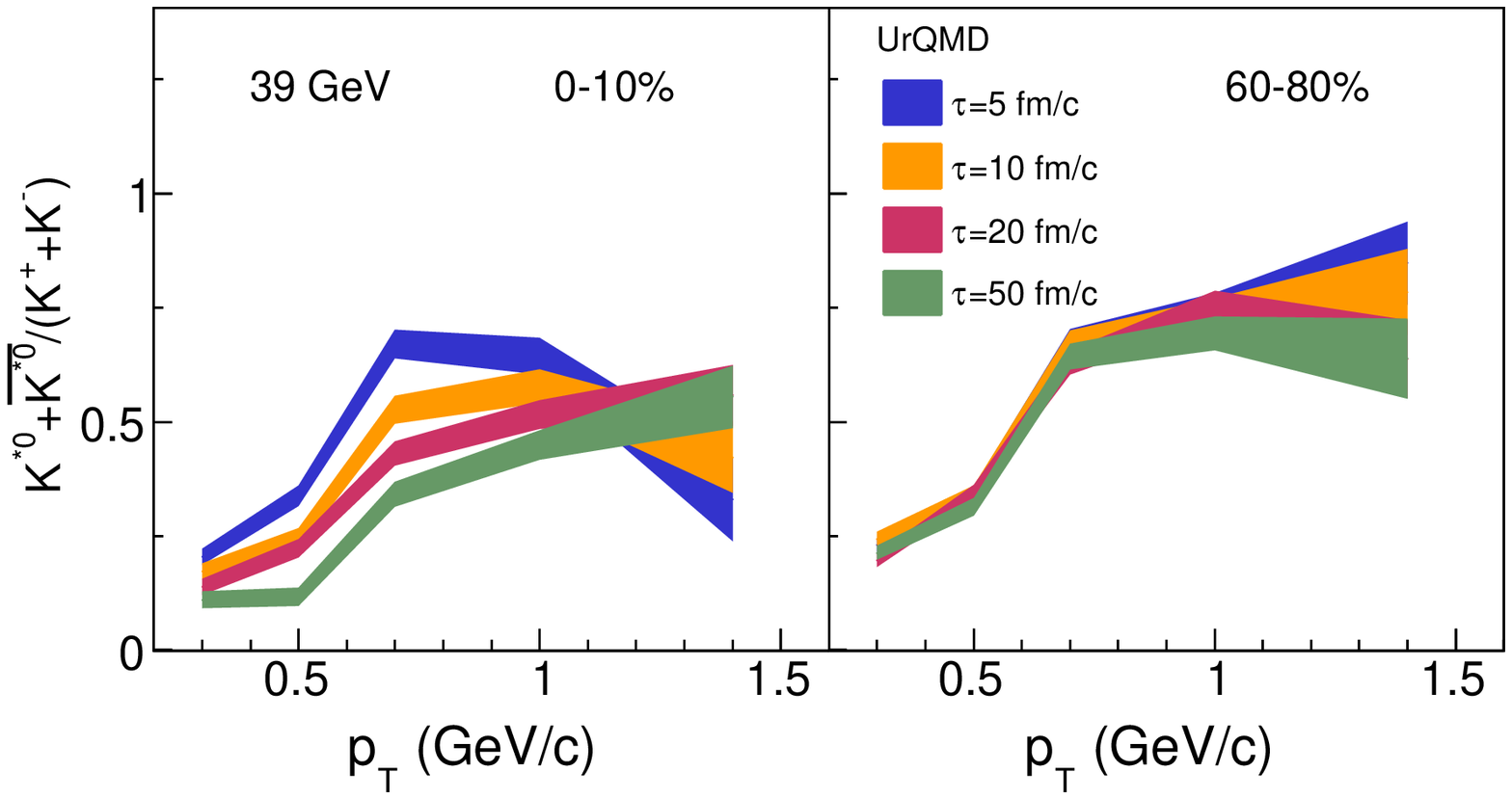}
\caption{(Color online) The $p_{T}$ dependence of $K^{*0}/K$ measured using the UrQMD model for 39 GeV at 0-10$\%$ centrality and 60-80$\%$ centrality.}
\label{yield_vspt_urqmd}
\end{center}
\eef

\subsection{Resonance to non-resonance ratio versus $N_{part}$ from UrQMD model, thermal model and STAR data}

The resonance to nonresonance ratios $(K^{*0}/K$ and $\phi/K)$ as a function of $N_{part}$ from the UrQMD model are shown in  Fig.~\ref{ratio_npart}, and compared with STAR data measured in Au+Au collisions at $\sqrt{s_{NN}}$ = 11.5 and 39 GeV.

The $K^{*0}/K$ ratios are found to decrease with increasing $N_{part}$. The $N_{part}$ dependence of $K^{*0}/K$ ratios is found to be similar to that measured by the STAR experiment for all STAR BES energies. The thermal model calculation are also shown in Fig.~\ref{ratio_npart}. The thermal model calculations for different $N_{part}$ are done by using different freeze-out parameters for corresponding centrality classes as mentioned in the Table~\ref{tab-freeze-out-param}. Note that there is no hadronic phase in the thermal model. Unlike UrQMD, the $K^{*0}/K$ ratio from the thermal model remains independent of $N_{part}$. The UrQMD measurements are done by varying $\tau$ from 5 to 50 fm/c. The $K^{*0}/K$ ratio at $\tau$= 5 fm/c remains almost independent of centrality, while a suppression can be observed for $\tau$= 10, 20, and 50 fm/c. For $\sqrt{s_{NN}}$= 39 GeV, the results from UrQMD with $\tau$= 50 fm/c seem to show better consistency with data compared to $\tau$= 20 fm/c. The data at $\sqrt{s_{NN}}$= 11.5 GeV can be explained by the UrQMD calculations with $\tau$= 20 fm/c.  UrQMD calculations at $\tau$= 20 and 50 fm/c give almost similar values of $K^{*0}/K$ ratio at 11.5 GeV, although a little difference is observed for central collisions at 39 GeV. If we consider the large statistical uncertainties the data at 11.5 GeV are also consistent with the UrQMD calculation for $\tau$= 10 fm/c. Hence measurement with higher statistics is needed to reach a precise conclusion. The high statistics data collected in the STAR beam energy scan phase-II program will help reduce the uncertainty in the measurement. As $\phi$ has nearly ten times longer lifetime than $K^{*0}$, one could naively expect it to decay outside the medium. Hence it would remain immune to the hadronic medium produced during the heavy-ion collisions. Hence $\phi/K$ remains almost independent of centrality and $\tau$. The trend is well explained by the thermal model, while UrQMD underpredicts the data~\cite{phi_steinheimer}.
The comparison of data with UrQMD and  the thermal model indicates that decay daughters of $K^{*0}$  suffer from late hadronic interaction, with rescattering playing a dominant role over regeneration, which is also evident from Fig.~~\ref{kstar_reco}.\\

 Furthermore, we conducted an analysis to extract the chemical freeze-out temperature using the hadron yield ratios obtained from the UrQMD model. The goal was to investigate whether there were any variations in the freezeout temperature as a function of centrality at a given $\tau$. Fig.~\ref{tch_npart} displays the centrality dependence of the chemical freezeout temperature ($t_{ch}$), obtained by fitting the thermal model to the UrQMD particle yield ratios ($\pi^{-}/\pi^{+}, K^{-}/K^{+}, K^{-}/\pi^{-}, K^{+}/\pi^{+},\,\bar{p}\,/\,p,\,p/\pi^{+}$).  A similar analysis for the chemical freeze-out temperature  using UrQMD data was conducted for HADES energies in~\cite{hades_thermal}. We found that $t_{ch}$ does not change appreciably with collision centrality. 
This observation further clarifies that the differences in the  $K^{*0}/K$ ratio between peripheral to central collisions is solely from (pseudo)elastic rescatterings.

\subsection{$K^{*0}/K$ versus transverse momentum ($p_T$) from UrQMD model}
The $K^{*0}/K$ ratio vs $p_{T}$ is measured from the UrQMD model is shown in Fig.~\ref{yield_vspt_urqmd} for central (0-10$\%$) and peripheral (60-80$\%$) Au+Au collisions at $\sqrt{s_{NN}}$ = 39 GeV, which could help detect the $p_{T}$ dependence of the rescattering effect. 

The $K^{*0}/K$ ratio is found to increase with $p_{T}$, which indicates that the low $p_{T}$ $K^{*0}$ mesons are more prone to undergo the rescattering effect than those at higher $p_{T}$. 
In the low $p_{T}$ region, $K^{*0}/K$ vs. $p_{T}$ weakly depends on the choice of $\tau$ in peripheral collisions compared to that in central collisions. A similar $p_{T}$ dependence was observed for $\sqrt{s_{NN}}$= 7.7-39 GeV. 

\bef
\begin{center}
\includegraphics[scale=0.7]{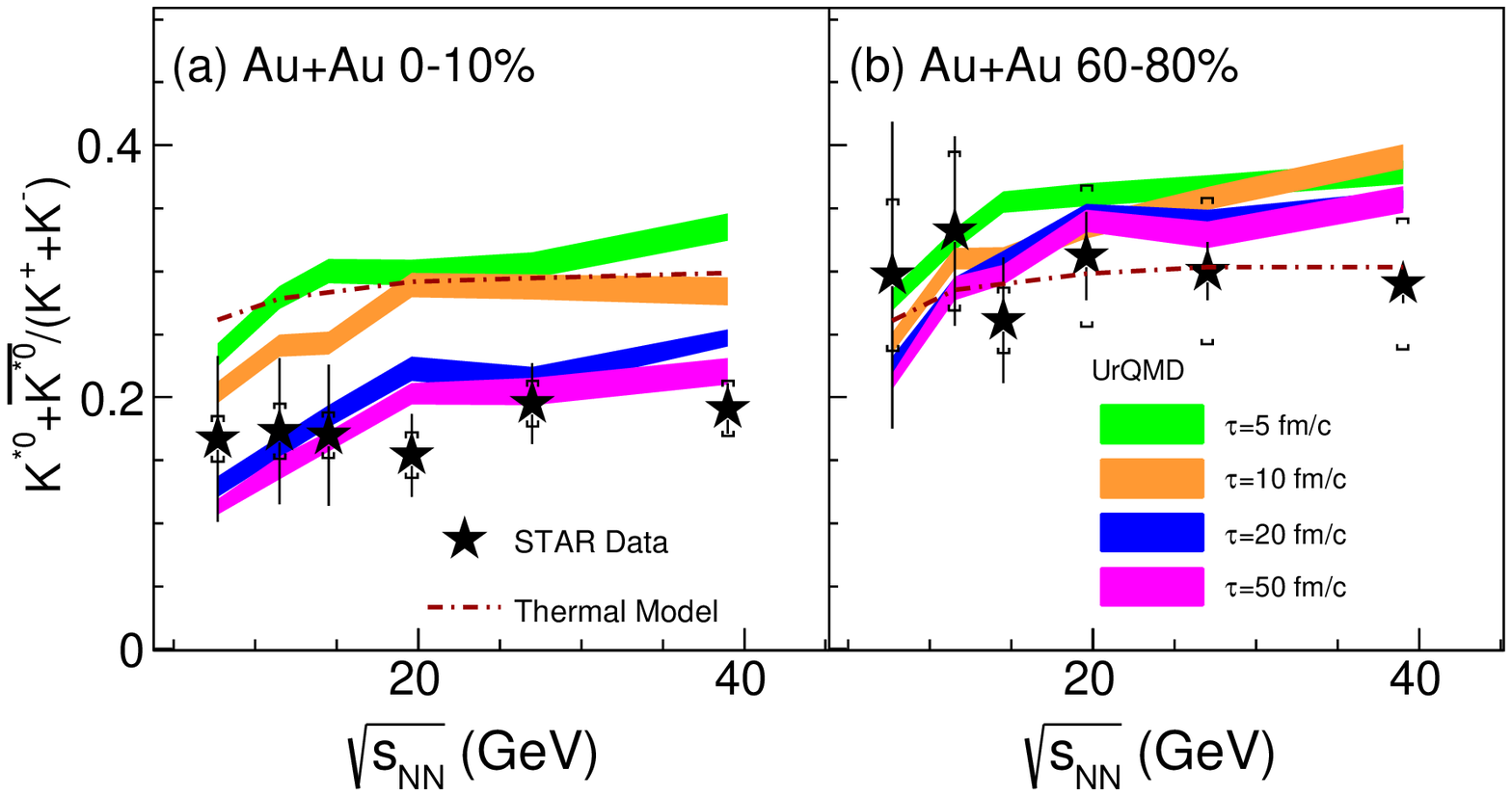}
\caption{(Color online) The $K^{*0}/K$ ratio vs center-of-mass energy for central (0-10$\%$) and peripheral (60-80$\%$) Au+Au collisions at midrapidity~\cite{kstar_BES} compared with corresponding measurements from thermal and UrQMD models. }
\label{ratio_snn}
\end{center}
\eef

\subsection{$K^{*0}/K$ versus $\sqrt{s_{NN}}$ (0-10$\%$ and 60-80$\%$) from UrQMD model and thermal model}
Figure~\ref{ratio_snn} shows energy dependence of $K^{*0}$/K for  0-10$\%$  central  and 60-80$\%$ peripheral Au+Au collisions. The STAR data do not show any significant energy dependence of $K^{*0}$/K for both 0-10$\%$  central  and 60-80$\%$ peripheral Au+Au collisions within present uncertainties. The UrQMD model calculation are shown for different $\tau$ values from 5 to 50 fm/c along with the thermal model prediction. \\
 The thermal model shows no centrality dependence. The overprediction of the data by the thermal model in central collisions is consistent with the expectation of dominance of hadronic rescattering. The $K^{*0}$/K ratio measured using the UrQMD model seems to increase with collision energy. However, a strong dependence on the hadronic cascade lifetime selection can be seen in central collisions as compared to the peripheral collisions. UrQMD results for  both $\tau$= 20 and 50 fm/c are consistent with the energy dependence of the $K^{*0}/K$ ratio at central collisions. The results below $\sqrt{s_{NN}}$= 14.5 GeV are also consistent with model prediction at $\tau$= 10 fm/c, within uncertainty. However the model results at peripheral collisions seem to be independent of $\tau$. This could indicate a smaller hadronic rescattering  at 60-80\% centrality as compared to 0-10\%.


\bef
\begin{center}
\includegraphics[scale=0.65]{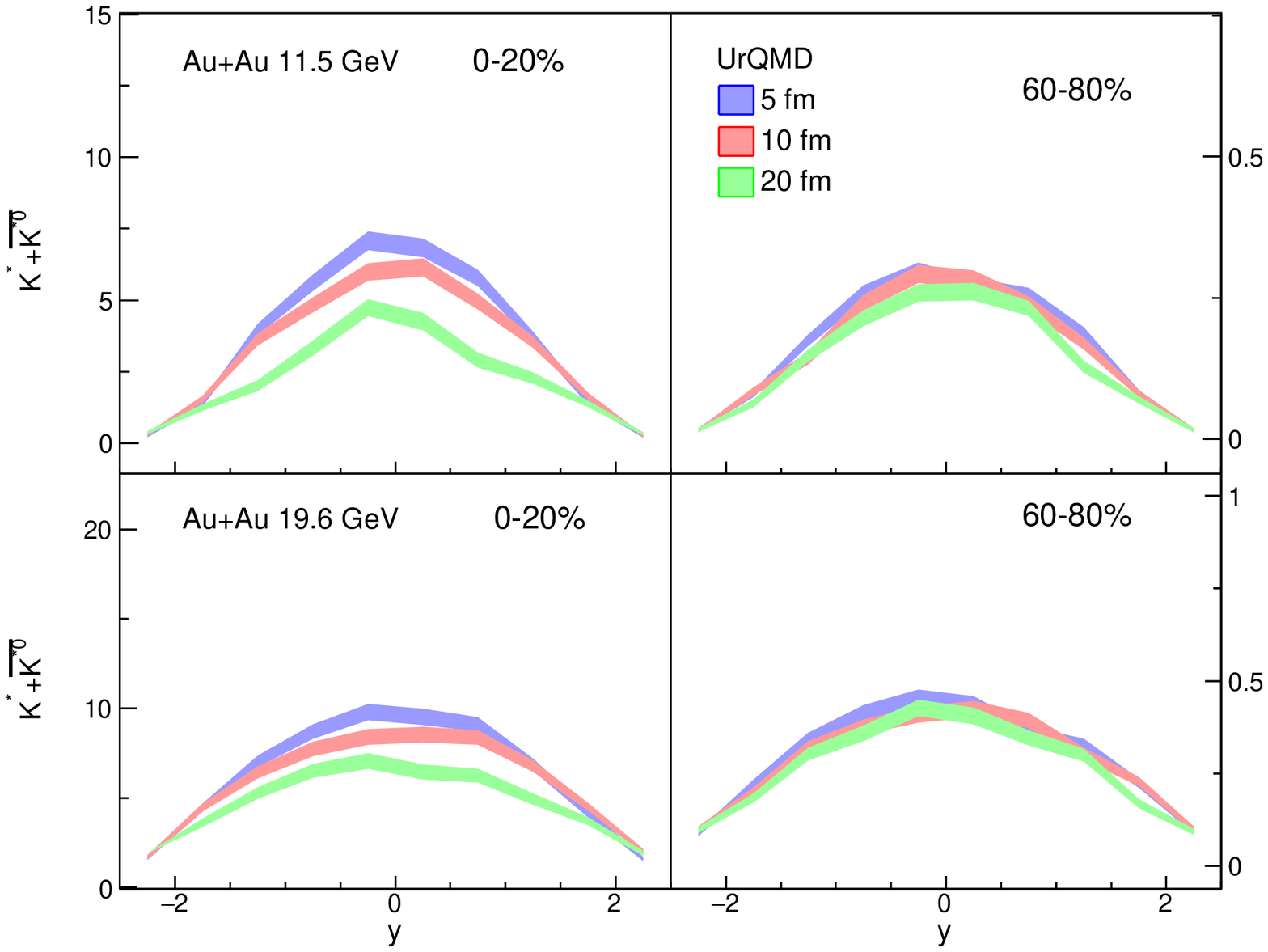}
\caption{(Color online) The $p_T$ integrated yield (dN/dy) for $K^{*0}$ mesons vs rapidity for 0-10\% and 60-80\% centrality at $\sqrt{s_{NN}}$= 11.5 GeV (upper panel) and 19.6 GeV (lower panel) respectively. }
\label{rap_dist}
\end{center}
\eef

\bef
\begin{center}
\includegraphics[scale=0.65]{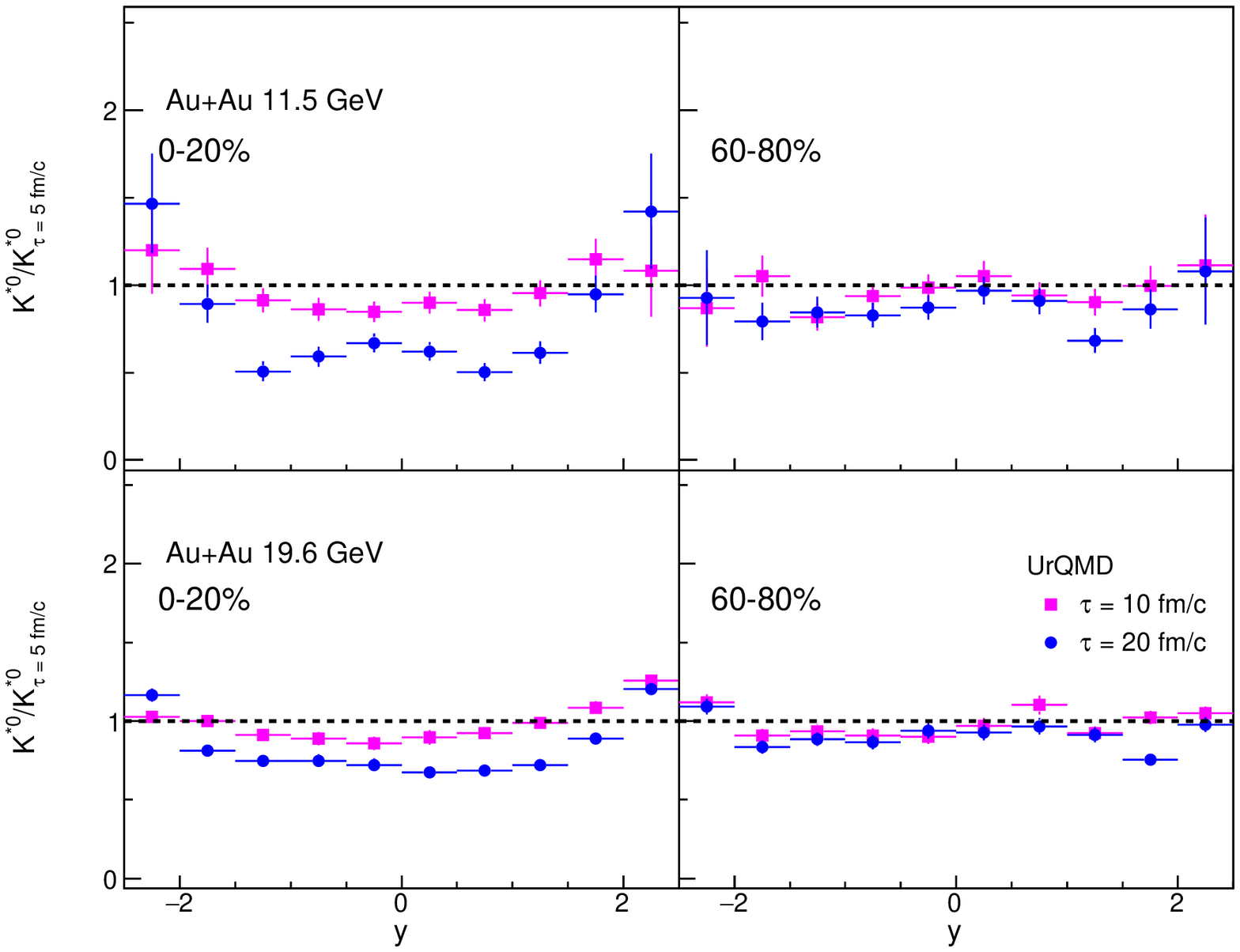}
\caption{(Color online) The $p_T$ integrated yield (dN/dy) for $K^{*0}$ mesons for $\tau$= 10 and 20 fm/c, divided by dN/dy for $\tau$= 5 fm/c as a function of rapidity for 0-10\% and 60-80\% centrality. }
\label{rap_dist_ratio}
\end{center}
\eef


\subsection{Rapidity dependence of $K^{*0}$ yield from UrQMD model}
The STAR experiment at RHIC has just finished data taking  for its phase II of the beam energy scan program. The data were taken with upgraded detectors providing an opportunity for measurement at a wider rapidity ($|y| <1.5$)~\cite{STAR_BESII_upgrade}. With high statistics data, the measurement for the $K^{*0}$ can be done as a function of rapidity to understand possible effect of hadronic rescattering when moving away from midrapidity.\\
In Fig~\ref{rap_dist}, the dN/dy of $K^{*0}$ meson is plotted as a function of rapidity for both central and peripheral collisions at at 11.5 and 19.6 GeV respectively. A clear rapidity dependance is observed for the $K^{*0}$ yield for all BES energies. However, a weaker dependence on $\tau$ is observed in peripheral collisions than in central collisions.\\
 In order to elucidate the effect of rescattering with rapidity ($y$), in Fig~\ref{rap_dist_ratio} the ratio of $K^{*0}$ yield (dN/dy) for $\tau$= 10, 20 fm/c is taken with that for $\tau$= 5 fm/c and plotted as a function of rapidity. Notably, for 19.6 GeV in central collisions, the ratio increases as we move towards larger rapidity. However, the ratios at 11.5 GeV do not exhibit such a smooth and monotonic change. The rescattering may play a more dominant role at midrapidity, where the particle density is higher. Furthermore, it can broaden the shape of rapidity distributions. Thus, any changes in the ratio as a function of rapidity arise from these effects.
For peripheral collisions the ratio remains almost independent of rapidity, which indicates that rescattering is not dominant in peripheral collisions.

\section{Summary}

We presented a comparison of the $K^{*0}$/K ratios measured at midrapidity in various centralities at RHIC BES energies with results of the UrQMD and thermal models. We found that $K^{*0}/K$ ratio from the thermal model, which does not include any hadronic rescattering, is consistent with data in most peripheral collisions but overpredicts the ratio in central Au+Au collisions. One needs to consider the hadronic rescattering in UrQMD to explain data at $\sqrt{s_{NN}}$ = 7.7-39 GeV in central Au+Au collisions. The UrQMD model calculations are done by changing the hadron propagation time. We found that the $K^{*0}/K$ ratio decreases with the increasing duration for hadronic interaction. This may suggest that the observed suppression of $K^{*0}/K$ ratio in central Au+Au collisions compared to peripheral collisions is due to the effect of hadronic rescattering suffered by the daughter particles of the $K^{*0}$ resonance. The study of the $\phi/K$ ratio from the UrQMD model further supports the idea of hadronic rescattering that the daughters of $K^{*0}$ resonance may undergo in central Au+Au collisions. In the end, we have made a prediction of the rapidity distribution of $K^{*0}$ yield using the UrQMD model. The study from the UrQMD model suggests that the rescattering is more dominant in the midrapidity region. These predicted values can be used to compare with STAR BES-II results to get more insight from the rapidity dependence study.\\

\noindent{\bf Acknowledgments}\\
A.K.S acknowledges discussions with Tribhuban Parida regarding thermal model calculations. SS acknowledges support from the Strategic Priority Research Program of the Chinese Academy of Sciences (Grant No. XDB34000000).



\begin{table*}
\begin{center}
\begin{tabular}{ |p{0.5cm}|p{0.5cm}|p{0.5cm}|p{0.5cm}|p{0.5cm}| }
\hline
\multicolumn{1}{|c}{$\sqrt{s_{NN}}$ (GeV)} & \multicolumn{1}{|c}{ Centrality ($\%$)} & \multicolumn{1}{|c}{$T_{ch}$ (MeV)} & \multicolumn{1}{|c}{$\mu_{B}$ (MeV)}& \multicolumn{1}{|c|}{$\mu_{s}$ (MeV)}\\
\hline
\multicolumn{1}{|c}{} & \multicolumn{1}{|c}{0-5} & \multicolumn{1}{|c}{144.3} & \multicolumn{1}{|c}{398.2}& \multicolumn{1}{|c|}{89.5}  \\
\multicolumn{1}{|c}{} & \multicolumn{1}{|c}{5-10} & \multicolumn{1}{|c}{143.0} & \multicolumn{1}{|c}{393.5}& \multicolumn{1}{|c|}{88.5}  \\
\multicolumn{1}{|c}{} & \multicolumn{1}{|c}{10-20} & \multicolumn{1}{|c}{143.8} & \multicolumn{1}{|c}{388.0}& \multicolumn{1}{|c|}{86.4}  \\
\multicolumn{1}{|c}{7.7} & \multicolumn{1}{|c}{20-30} & \multicolumn{1}{|c}{143.5} & \multicolumn{1}{|c}{379.5}& \multicolumn{1}{|c|}{85.2}  \\
\multicolumn{1}{|c}{} & \multicolumn{1}{|c}{30-40} & \multicolumn{1}{|c}{145.9} & \multicolumn{1}{|c}{375.4}& \multicolumn{1}{|c|}{85.5}  \\
\multicolumn{1}{|c}{} & \multicolumn{1}{|c}{40-60} & \multicolumn{1}{|c}{144.7} & \multicolumn{1}{|c}{355.6}& \multicolumn{1}{|c|}{80.3}  \\
\multicolumn{1}{|c}{} & \multicolumn{1}{|c}{60-80} & \multicolumn{1}{|c}{143.4} & \multicolumn{1}{|c}{337.5}& \multicolumn{1}{|c|}{79.3}  \\
\hline
\multicolumn{1}{|c}{} & \multicolumn{1}{|c}{0-5} & \multicolumn{1}{|c}{149.4} & \multicolumn{1}{|c}{287.3}& \multicolumn{1}{|c|}{64.5}  \\
\multicolumn{1}{|c}{} & \multicolumn{1}{|c}{5-10} & \multicolumn{1}{|c}{150.1} & \multicolumn{1}{|c}{288.9}& \multicolumn{1}{|c|}{65.8}  \\
\multicolumn{1}{|c}{} & \multicolumn{1}{|c}{10-20} & \multicolumn{1}{|c}{151.8} & \multicolumn{1}{|c}{284.9}& \multicolumn{1}{|c|}{65.1}  \\
\multicolumn{1}{|c}{11.5} & \multicolumn{1}{|c}{20-30} & \multicolumn{1}{|c}{153.5} & \multicolumn{1}{|c}{278.7}& \multicolumn{1}{|c|}{63.9}  \\
\multicolumn{1}{|c}{} & \multicolumn{1}{|c}{30-40} & \multicolumn{1}{|c}{154.6} & \multicolumn{1}{|c}{270.1}& \multicolumn{1}{|c|}{61.9}  \\
\multicolumn{1}{|c}{} & \multicolumn{1}{|c}{40-60} & \multicolumn{1}{|c}{155.3} & \multicolumn{1}{|c}{256.0}& \multicolumn{1}{|c|}{60.2}  \\
\multicolumn{1}{|c}{} & \multicolumn{1}{|c}{60-80} & \multicolumn{1}{|c}{151.6} & \multicolumn{1}{|c}{227.3}& \multicolumn{1}{|c|}{54.6}  \\
\hline
\multicolumn{1}{|c}{} & \multicolumn{1}{|c}{0-5} & \multicolumn{1}{|c}{153.9} & \multicolumn{1}{|c}{187.9}& \multicolumn{1}{|c|}{43.2}  \\
\multicolumn{1}{|c}{} & \multicolumn{1}{|c}{5-10} & \multicolumn{1}{|c}{154.2} & \multicolumn{1}{|c}{187.2}& \multicolumn{1}{|c|}{43.9}  \\
\multicolumn{1}{|c}{} & \multicolumn{1}{|c}{10-20} & \multicolumn{1}{|c}{155.9} & \multicolumn{1}{|c}{184.9}& \multicolumn{1}{|c|}{44.4}  \\
\multicolumn{1}{|c}{19.6} & \multicolumn{1}{|c}{20-30} & \multicolumn{1}{|c}{156.4} & \multicolumn{1}{|c}{177.2}& \multicolumn{1}{|c|}{42.6}  \\
\multicolumn{1}{|c}{} & \multicolumn{1}{|c}{30-40} & \multicolumn{1}{|c}{157.5} & \multicolumn{1}{|c}{166.9}& \multicolumn{1}{|c|}{40.3}  \\
\multicolumn{1}{|c}{} & \multicolumn{1}{|c}{40-60} & \multicolumn{1}{|c}{157.9} & \multicolumn{1}{|c}{154.4}& \multicolumn{1}{|c|}{38.0}  \\
\multicolumn{1}{|c}{} & \multicolumn{1}{|c}{60-80} & \multicolumn{1}{|c}{156.2} & \multicolumn{1}{|c}{133.7}& \multicolumn{1}{|c|}{33.3}  \\
\hline
\multicolumn{1}{|c}{} & \multicolumn{1}{|c}{0-5} & \multicolumn{1}{|c}{155.0} & \multicolumn{1}{|c}{144.4}& \multicolumn{1}{|c|}{33.5}  \\
\multicolumn{1}{|c}{} & \multicolumn{1}{|c}{5-10} & \multicolumn{1}{|c}{155.6} & \multicolumn{1}{|c}{143.9}& \multicolumn{1}{|c|}{34.1}  \\
\multicolumn{1}{|c}{} & \multicolumn{1}{|c}{10-20} & \multicolumn{1}{|c}{155.8} & \multicolumn{1}{|c}{137.7}& \multicolumn{1}{|c|}{32.0}  \\
\multicolumn{1}{|c}{27} & \multicolumn{1}{|c}{20-30} & \multicolumn{1}{|c}{157.1} & \multicolumn{1}{|c}{131.0}& \multicolumn{1}{|c|}{31.0}  \\
\multicolumn{1}{|c}{} & \multicolumn{1}{|c}{30-40} & \multicolumn{1}{|c}{158.9} & \multicolumn{1}{|c}{130.3}& \multicolumn{1}{|c|}{32.4}  \\
\multicolumn{1}{|c}{} & \multicolumn{1}{|c}{40-60} & \multicolumn{1}{|c}{160.4} & \multicolumn{1}{|c}{120.4}& \multicolumn{1}{|c|}{31.4}  \\
\multicolumn{1}{|c}{} & \multicolumn{1}{|c}{60-80} & \multicolumn{1}{|c}{158.3} & \multicolumn{1}{|c}{105.8}& \multicolumn{1}{|c|}{28.6}  \\
\hline
\multicolumn{1}{|c}{} & \multicolumn{1}{|c}{0-5} & \multicolumn{1}{|c}{156.4} & \multicolumn{1}{|c}{103.2}& \multicolumn{1}{|c|}{24.5}  \\
\multicolumn{1}{|c}{} & \multicolumn{1}{|c}{5-10} & \multicolumn{1}{|c}{157.0} & \multicolumn{1}{|c}{101.9}& \multicolumn{1}{|c|}{24.8}  \\
\multicolumn{1}{|c}{} & \multicolumn{1}{|c}{10-20} & \multicolumn{1}{|c}{156.3} & \multicolumn{1}{|c}{101.9}& \multicolumn{1}{|c|}{24.9}  \\
\multicolumn{1}{|c}{39} & \multicolumn{1}{|c}{20-30} & \multicolumn{1}{|c}{157.9} & \multicolumn{1}{|c}{98.2}& \multicolumn{1}{|c|}{24.9}  \\
\multicolumn{1}{|c}{} & \multicolumn{1}{|c}{30-40} & \multicolumn{1}{|c}{160.8} & \multicolumn{1}{|c}{94.2}& \multicolumn{1}{|c|}{24.0}  \\
\multicolumn{1}{|c}{} & \multicolumn{1}{|c}{40-60} & \multicolumn{1}{|c}{160.0} & \multicolumn{1}{|c}{84.6}& \multicolumn{1}{|c|}{21.9}  \\
\multicolumn{1}{|c}{} & \multicolumn{1}{|c}{60-80} & \multicolumn{1}{|c}{158.3} & \multicolumn{1}{|c}{73.0}& \multicolumn{1}{|c|}{20.3}  \\

\hline
\end{tabular}
\end{center}
\caption{Freeze-out parameters at different centralities in Au+Au collisions at $\sqrt{s_{NN}}$ = 7.7, 11.5, 19.6, 27, 39 GeV, taken from~\cite{bulk_BES} }
\label{tab-freeze-out-param}
\end{table*}

\end{document}